\documentclass[prb,twocolumn,superscriptaddress,showpacs,preprintnumbers,amsmath,amssymb,floatfix]{revtex4}
\usepackage{mathbbol}
\usepackage{graphicx}
\usepackage{epstopdf}
\usepackage{dcolumn}
\usepackage{bm}
\usepackage{array}
\usepackage{hyperref}

\begin{document}
	\bibliographystyle{plain}
	\newcommand*{\cm}{$cm^{-1}$}
	\newcommand*{\Tn}{$T_N$}
	\newcommand*{\sample}{$\mathrm{EuCd_2As_2}$}
	\newcommand*{\R}{${R(\omega)}$}
	\newcommand*{\s}{${\sigma_1(\omega)}$}
	\newcommand*{\omegap}{${\omega_p}$}
	
	\title{Anisotropic transport and optical spectroscopy study on antiferromagentic triangular lattice $\mathrm{EuCd_2As_2}$: an interplay between magnetism and charge transport properties}
	
	\author{H. P. Wang}
	\affiliation{Beijing National Laboratory for Condensed Matter Physics, Institute of Physics, Chinese Academy of Sciences, Beijing 100190, China}
	
	\author{D. S. Wu}
	\author{Y. G. Shi}
	\affiliation{Beijing National Laboratory for Condensed Matter Physics, Institute of Physics, Chinese Academy of Sciences, Beijing 100190, China}
	
	\author{N. L. Wang}
	\email{nlwang@pku.edu.cn}
	\affiliation{International Center for Quantum Materials, School of Physics, Peking University, Beijing 100871, China}
	\affiliation{Collaborative Innovation Center of Quantum Matter, Beijing 100871, China}
	
	\begin{abstract}
We present anisotropic transport and optical spectroscopy studies on $\mathrm{EuCd_2As_2}$. The measurements reveal that $\mathrm{EuCd_2As_2}$ is a low carrier density semimetal with moderate anisotropic resistivity ratio. The charge carriers experience very strong scattering from Eu magnetic moments, resulting in a Kondo-like increase of resistivity at low temperature. Below the antiferromagnetic transition temperature at $T_N$= 9.5 K, the resistivity drops sharply due to the reduced scattering from the ordered Eu moments. Nevertheless, the anisotropic ratio of $\rho_c/\rho_{ab}$ keeps increasing, suggesting that the antiferromagnetic coupling is along the c-axis. The optical spectroscopy measurement further reveals, besides an overdamped reflectance plasma edge at low energy, a strong coupling between phonon and electronic continuum. Our study suggests that $\mathrm{EuCd_2As_2}$ is a promising candidate displaying intriguing interplay among charge, magnetism and the underlying crystal lattice.
	\end{abstract}
	
	 \maketitle
	
	\section{Introduction}
	Low-dimensional magnetic systems have been a subject of considerable theoretical and experimental studies in condensed matter physics. A large variety of interesting physical phenomena, such as high temperature superconductivity\cite{Bednorz,Schilling1993,Hosono2008} and colossal magnetoresistance\cite{Han2010}, have been observed in such systems. Due to the complex coupling among charge, spin and lattice degrees of freedoms in those systems, a subtle change of an external parameter (e.g. temperature, magnetic field or pressure) may result in a dramatic change of the physical properties. Among different low-dimensional electronic systems, the so-called $\mathrm{AB_2X_2}$ (122) compounds with $\mathrm{ThCr_2Si_2}$\cite{Ban1965,Just1996} structure have been extensively studied since the discovery of Fe-based superconductors in K-doped $\mathrm{BaFe_2As_2}$\cite{Rotter2008}. The structure is also well known to host many other intriguing quantum electronic states, such as heavy fermions\cite{Steglich1997,Palstra1985} and quantum criticality\cite{Gegenwart2008,Xu2008,Nakai2010}. $\mathrm{ThCr_2Si_2}$ has the tetragonal structure and belongs to the I4/nmm space group. In this structure, the  $B_2X_2$ layers, which are separated by A site ions, are composed of an edge-sharing network of $BX_4$ tetrahedra.
		
	\begin{figure}
		\centering
		\includegraphics[width=8cm]{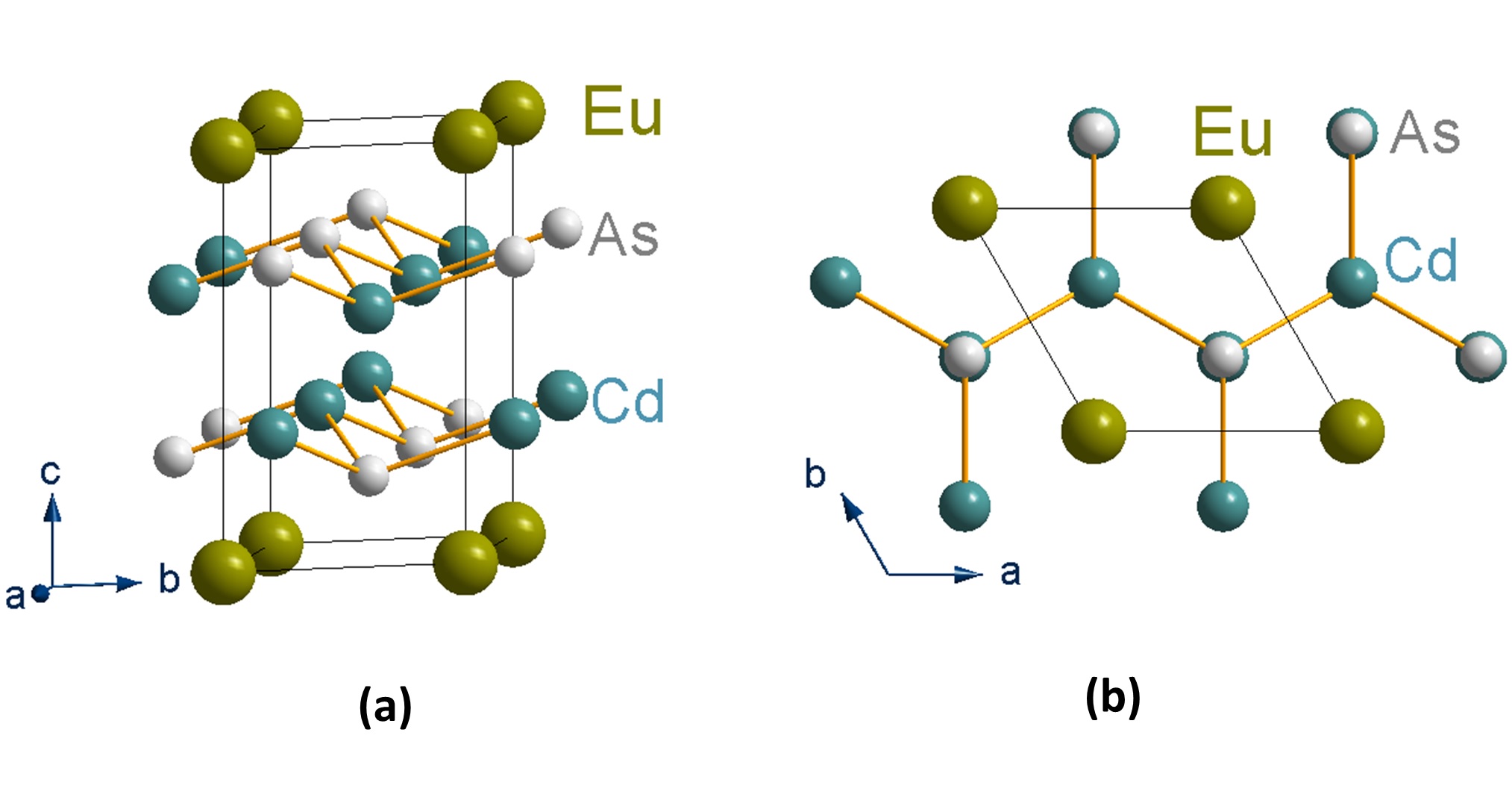}\\
		\caption{(a) The crystal structure of $\mathrm{EuCd_2As_2}$. (b) The crystal structure viewed from the c-axis. Cations within each layer (ab-plane) have a triangle lattice.}
		\label{Fig:Struc}
	\end{figure}
	
	$\mathrm{AB_2X_2}$ has a related but much less studied $\mathrm{CaAl_2Si_2}$-type structure\cite{Bodak1967}. This is also a layered structure with the $B_2X_2$-layers separated by A site ions. Different from $\mathrm{ThCr_2Si_2}$,  $\mathrm{CaAl_2Si_2}$ has a trigonal structure with a $P\overline{3}m1$ space group. Cations within each layer (ab-plane) have a triangle lattice, as shown in Fig.~\ref{Fig:Struc}. Prominent examples are the prototype itself $\mathrm{CaAl_2Si_2}$ or oxide sulfides like $\mathrm{Ce_2O_2S}$\cite{Zachariasen1948}. The latter compounds have widely been investigated as luminescent materials. In recent years, several europium and ytterbium containing ternary pnictides with $\mathrm{CaAl_2Si_2}$-type structure have been synthesized and studied due to their promising magnetic or thermoelectric properties\cite{Zhang2010,MIN2015,Goryunov2012,Schellenberg2011}. In particular, a recent study on the \sample~compound revealed an antiferromagnetic order with $T_N$=9.5 K in such a triangular lattice \cite{Schellenberg2011}. The effective magnetic moment derived from magnetization measurement is 7.88 $\mu_B$, which is very close to the theoretical value of 7.94 $\mu_B$ for a free Eu$^{2+}$ ion \cite{Schellenberg2011}. Therefore, the magnetic moments come from the Eu sites with an electronic configuration of $4f^7$. Notably, although the compound shows an antiferromagnetic order at low temperature, a positive Curie-Weiss temperature was obtained from the study of magnetic susceptibility and the M\"{o}ssbauer spectroscopy \cite{Goryunov2012,Schellenberg2011}, indicating the presence of ferromagnetic fluctuation above $T_N$. At present, little is known about its physical properties, except for the magnetic susceptibility and Eu magnetic moments. Since Cd has an electron configuration of $4d^{10}5s^2$  and As has a configuration of $4s^24p^3$, the compound could be either a semiconductor or a semimetal depending on the ionic states of Cd and As. Assuming that Cd and As elements in $\mathrm{EuCd_2As_2}$ have ideal ionic states of Cd$^{2+}$ and As$^{3-}$, meaning that the 5s orbital of Cd is completely empty and the 4p orbital of As is completely filled, then the compound would be a semiconductor. On the other hand, if the Cd and As do not have the above ideal valence states, that is, 5s orbital of Cd is not completely empty and the 4p orbital of As is not fully filled, the compound would be a semimetal with small and compensated electron and hole carrier densities. Thus it would be of high interest to investigate its electronic properties and to elaborate possible interplay between charge transport and the magnetism formed by the Eu sublattice.
	
	In this work we present anisotropic transport and optical spectroscopy studies on $\mathrm{EuCd_2As_2}$. The compound shows metallic transport property with moderate anisotropic ratio. At low temperature, the conducting carriers suffer strong scattering from the Eu magnetic moments, resulting in a Kondo-like increase of resistivity. Below $T_N$, the resistivity drops sharply due to the reduced scattering from the ordered state of Eu moments. Nevertheless, the anisotropic ratio of $\rho_c/\rho_{ab}$ keeps increasing, suggesting that the antiferromagnetic coupling is along the c-axis. We also identified a very large negative magnetoresistance at low temperature. The optical spectroscopy measurement reveals that $\mathrm{EuCd_2As_2}$ has a small plasma frequency with an overdamped shape of reflectance edge, yielding further evidence that the compound is a low carrier density semimetal with very strong scattering effect. In the far-infrared spectra, we observed two infrared-active phonons superimposed in the electronic background. One of them shows a strong asymmetric line-shape, indicating a strong coupling between phonon and electronic continuum. Our measurements suggest that $\mathrm{EuCd_2As_2}$ is a promising candidate displaying intriguing interplay between magnetism, charge and the underlying crystal lattice.

\section{Experiments}

\begin{figure}
		\includegraphics[width=7 cm]{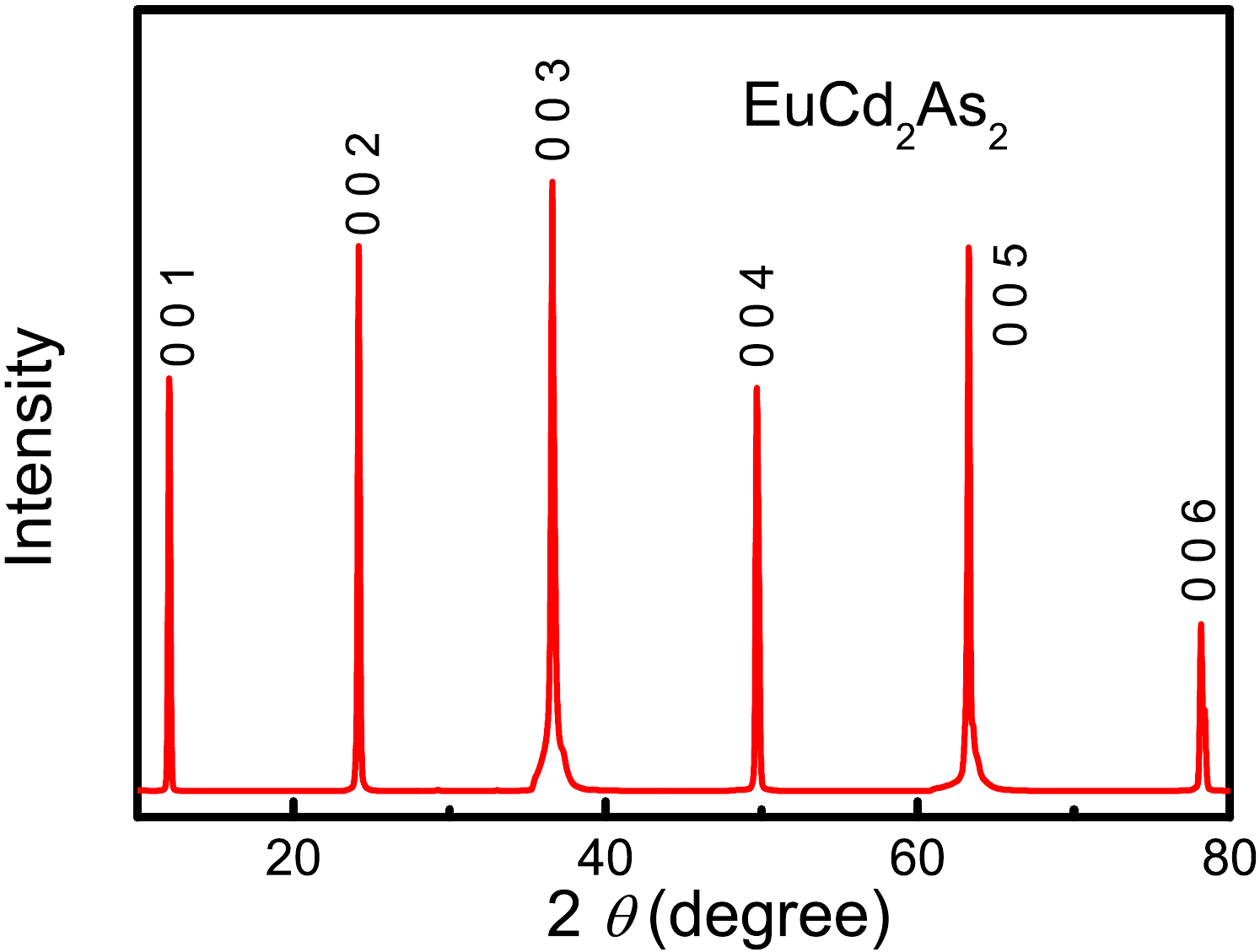}
		\caption{(Color online) C-axis X-ray diffraction pattern of $\mathrm{EuCd_2As_2}$ single crystal. The extracted c-axis lattice parameter is 7.34 {\AA}. \label{fig:XRD}}
	\end{figure}

	The $\mathrm{EuCd_2As_2}$ single crystals were grown using the flux method as described in an earlier report \cite{Schellenberg2011}. The obtained crystals have triangular and plate-like shape. They were characterized by both powder and single crystal X-ray diffractions, and pure $\mathrm{CaAl_2Si_2}$ structure phase was confirmed. Figure \ref{fig:XRD} shows the c-axis X-ray diffraction pattern for a crystal. Only sharp (0 0 l) diffraction lines are present. The in-plane resistivity $\rho_{ab}$ and out-of-plane resistivity $\rho_{c}$ were measured and calculated by the Montgomery method\cite{Montgomery1971,Logan1971}. The optical reflectance measurements were performed on a combination of Bruker IFS 80v/s and 113 v spectrometers in the frequency range from 30 to 40000 \cm. An \emph{in situ} gold and aluminum overcoating technique was used to get the reflectivity R($\omega$). The real part of conductivity $\sigma_1(\omega)$ is obtained by the Kramers-Kronig transformation of R($\omega$) employing an extrapolation method with X-ray atomic scattering functions\cite{Tanner2015}. This new Kramers-Kronig transformation method is proved to be more effective and unambiguous in deriving and analyzing the optical constants.

\section{Anisotropic resistivity}

	\begin{figure}
		\includegraphics[width=8.5cm]{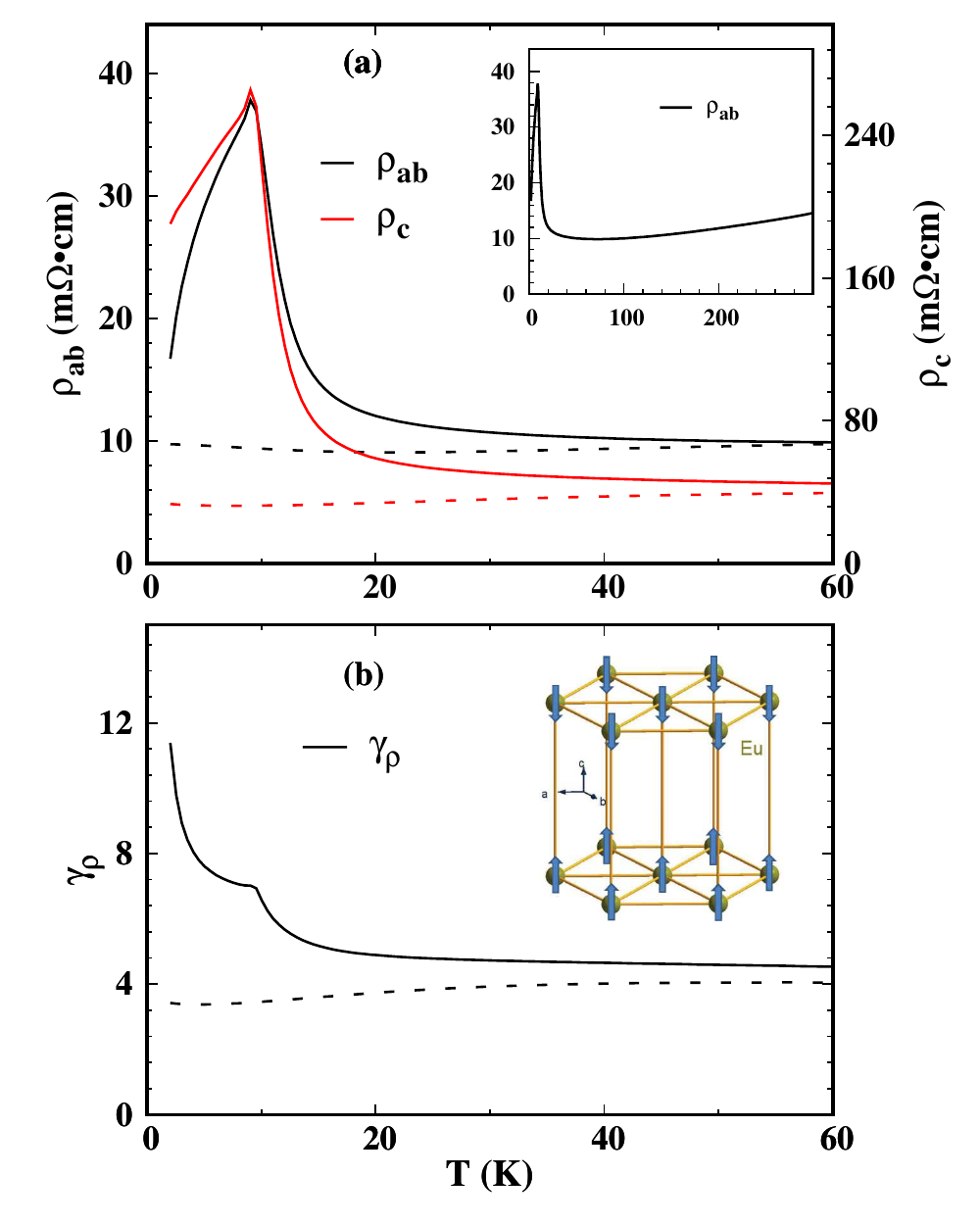}
		\caption{(Color online) (a) Temperature dependent in-plane resistivity $\rho_{ab}$ and $\rho_{c}$ at range from 0 to 60 K under zero field (solid lines) and 9 T (dot lines). The inset shows $\rho_{ab}$ from 0 to 300 T. (b) Temperature dependent resistivity anisotropy $\gamma_\rho=\rho_{c}/\rho_{ab}$ at range from 0 to 60 K under zero field (solid lines) and 9 T (dot lines). The inset displays a proposed A-type antiferromagnetic structure on Eu sites. \label{fig:RT}}
	\end{figure}

	Figure \ref{fig:RT} (a) shows the temperature dependent in-plane and out-of-plane resistivity $\rho_{ab}$ and $\rho_{c}$. At high temperature, the in-plane resistivity $\rho_{ab}$ has a positive slope, showing typical metallic behavior.
On the contrary, at low temperature $\rho_{ab}$ increases dramatically to a maximum at $ T_N $ = 9.5 K. The out-of-plane resistivity $ \rho_c $ nearly shows the same behavior as $\rho_{ab}$ and a moderate anisotropic ratio ($4\sim 6$) is observed above $T_N$. The Kondo-like increase of resistivity should be ascribed to the strong scattering of conducting carriers from the Eu magnetic moments. The overall metallic transport suggests the presence of free carrier contribution to the conductivity. Considering the valence balance or charge neutrality, the compound should be a semimetal with small and compensated electron and hole carrier densities. The electron pocket should come from a Cd 5s orbital since it is not completely empty, while the hole pocket should originate from an As 4p orbital as it is not fully filled. The conclusion is supported by optical spectroscopy measurement, as we shall present in the next section.

Below AFM transition $T_N$, both $\rho_{ab}$ and $\rho_c$ drop sharply. This is due to the reduced scattering from the ordered state of Eu moments. Nevertheless, the anisotropic ratio of $\rho_c/\rho_{ab}$ keeps increasing, as seen in Fig. \ref{fig:RT} (b). The observation suggests that the carriers experience stronger scattering in the c-axis in the ordered state. The result can be easily understood if we assume that the Eu moments are parallel to the c-axis, and are ferromagnetic ordered within the ab-plane but antiferromagnetic coupled along the c-axis. On this basis, the compound should have a so-called A-type antiferromagnetic structure in the ordered state, as displayed in the inset of Fig. \ref{fig:RT} (b). Of course, the true magnetic structure should be determined by future neutron diffraction measurement.

We also performed anisotropic resistivity measurement on the sample under high magnetic field. The in-plane resitivity $\rho_{ab}$, the c-axis resistivity $\rho_c$, as well as the anisotropic ratio $\gamma_\rho=\rho_{c}/\rho_{ab}$ under a magnetic field of 9 T with $\textbf{B}\parallel$c-axis are shown as dashed lines in Fig. \ref{fig:RT} (a) and (b), respectively. The resistivity has lower values relative to the zero magnetic field, \emph{i.e.} a negative magnetoresistance is present. In particular at low temperature, e. g. near 9 K, the drop of resistance is up to 80\%, which implies a giant negative magnetoresistance effect. The effect is similar to that observed in $\mathrm{TiTeI}$\cite{Guo2014} which shares the same space group with $\mathrm{EuCd_2As_2}$. Furthermore, no anomaly associated with the AFM transition $T_N$ is observed. As it is known from the magnetic susceptibility measurement, the magnetization from Eu moments is saturated when the field is higher than 1 T \cite{Schellenberg2011}. In our measurement at a field of 9 T, all the Eu moments should be aligned along the field direction (c-axis), the conducting carriers experience scattering from ferromagnetic ordered Eu moments in both directions, therefore no peak structure is present and the anisotropic ratio is roughly in a range of  $\rho_c/\rho_{ab}$=3$\sim$4.

\section{In-plane optical response}

	\begin{figure}
		\includegraphics[width=8cm]{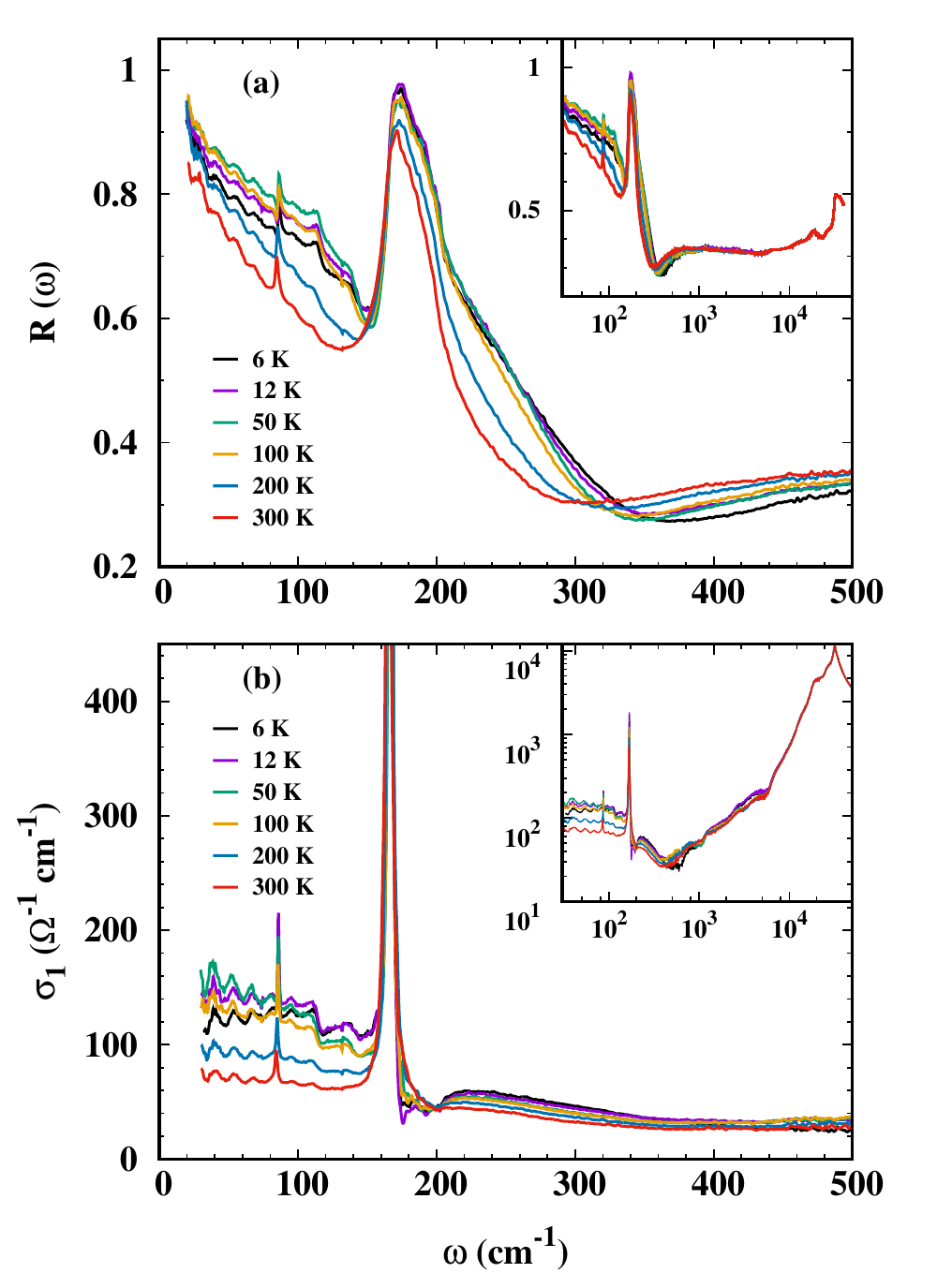}
		\caption{(Color online) (a) Optical reflectivity \R~between the frequency 20 \cm~and 40000 \cm~at six representative temperature. The inset show a broad range using logscale for x axis. (b) Optical conductivity \s~between the frequency 30 \cm~and 500 \cm~at six representative temperature. The inset: A broad view up to 40000 \cm~ using logscale for x and y axes. \label{fig:R-sigma1}}
	\end{figure}

	Figure \ref{fig:R-sigma1} shows the temperature dependent in-plane optical reflectivity {$R(\omega)$} (a) and real part of optical conductivity $\sigma_1(\omega)$ (b) of $\mathrm{EuCd_2As_2}$. The reflectivity $R(\omega)$ shows a plasma edge below 350 \cm~and approaches unity at low frequency. Both the temperature and frequency dependent optical response indicates clearly a metallic response. The edge frequency, also referred to as the "screened" plasma frequency, is related to the carrier density $n$ and effective mass $m^*$ of free carriers by $\omega_p'^2 \propto n/m^*$. The unusual low value of this frequency suggests a very small conducting carrier density. As temperature decreases, the edge frequency moves slightly to higher energy. We notice that $R(\omega)$ decreases continuously with frequency, so that the shape of the plasma edge is not well-defined. Such over-damped behavior also reflects the fact that the charge carriers experience very strong scattering from the magnetic moments. At 6 K and 12 K, $R(\omega)$ of the low frequency range are suppressed to lower values. As a consequence, a weak suppression is also seen in $\sigma_1(\omega)$ at low frequencies. The behavior of \s~at low frequency range slightly differs from the Drude response of free carriers. This difference is also likely to be caused by the strong scattering of magnetic moments at low temperature which makes free carriers prone to be localized. The behavior is similar to the effect of weak Anderson localization but the underlying mechanism is different. The magnetic scattering, or Kondo physics, should be dominant physical reason. In the frequency range from 1000 \cm~to 10000 \cm~, the $R(\omega)$ is nearly independent of temperature and frequency [inset of Fig. \ref{fig:R-sigma1}(a)]. It is worth mentioning that the oscillations below 100 \cm~are likely due to the interference between light beams from cracked layers of the sample and do not influence the main results we discuss here.

	\begin{figure}
		\includegraphics[width=8cm]{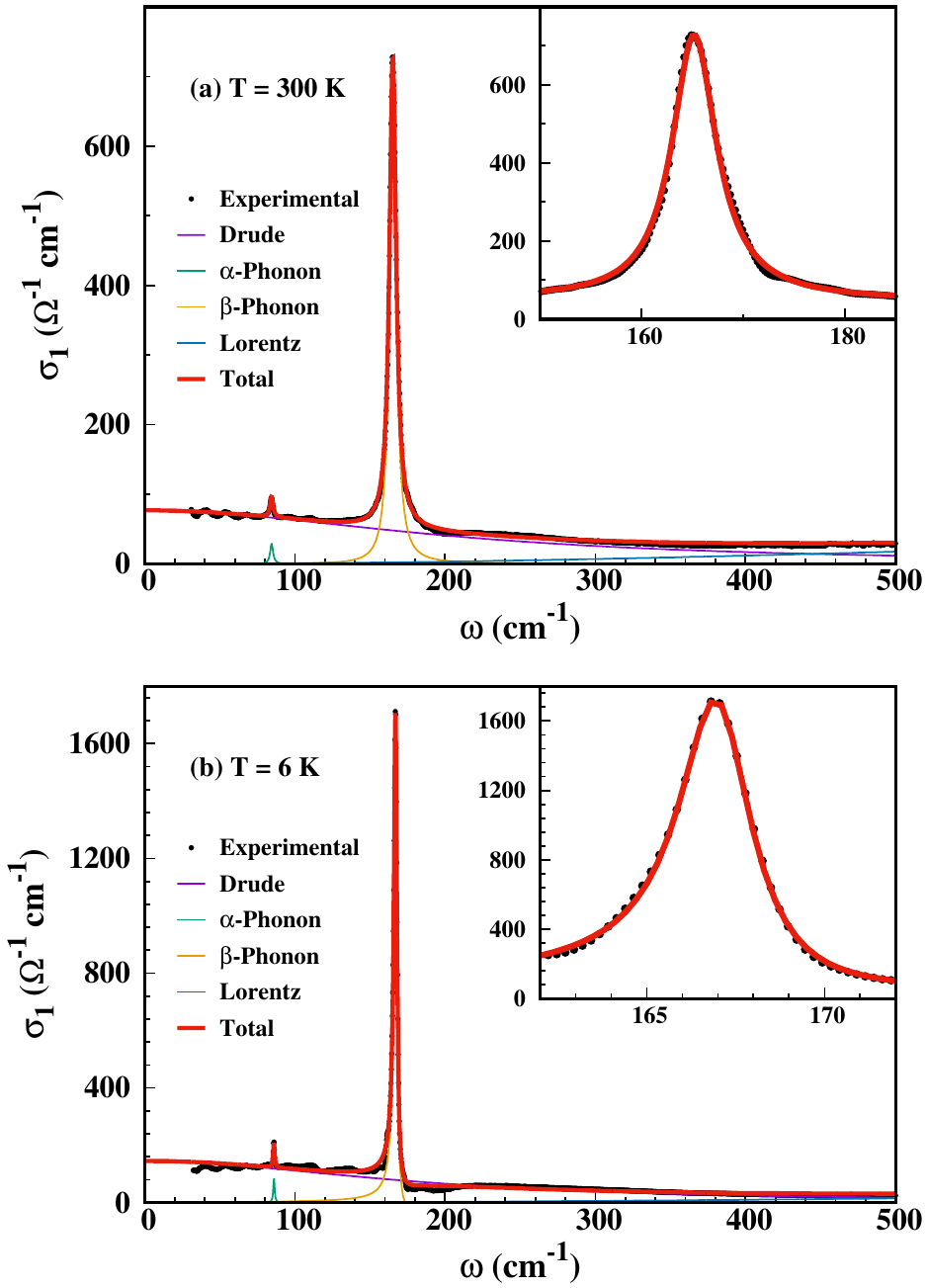}
		\caption{(Color online) Optical conductivity \s~together with their fitting results at 300 K (a) and 6 K (b). The insets: experimental and fitting results of the $\beta$ phonon.\label{fig:fit}}
	\end{figure}
	
		\begin {table*}
		\caption {Fit Parameters.}
		\label {FitTable}
		\centering
		\begin {tabular}{p{1.25cm}<{\centering} | p{1.25cm}<{\centering}p{1.25cm}<{\centering} | p{1.25cm}<{\centering}p{1.25cm}<{\centering}p{1.25cm}<{\centering}p{1.25cm}<{\centering}|p{1.25cm}<{\centering}p{1.25cm}<{\centering}p{1.25cm}<{\centering}}
		\toprule %
		
		& \multicolumn{2}{c|}{Drude} & \multicolumn{4}{c|}{$\beta$-phonon(Fano)} & \multicolumn{3}{c}{$\alpha$-phonon}\\
		\hline
		T & $\omega_p$ &  $\Gamma$  &$\omega_{0}$ &  $\Gamma$ & $\omega_{p,Fano}$ & $\Theta$ &$\omega_{0}$ &  $\Gamma_{}$ & $\omega_{p0}$  \\
		(K) &(\cm)&  (\cm) & (\cm) & (\cm) & (\cm) &   & (\cm) & (\cm) & (\cm) \\
		\hline
		300 & 900 	& 190 	& 165.2 & 4.37 & 431 & 0		& 84.6 & 2.4 & 64.1\\
		200 & 1090  & 194 	& 166.2 & 4.18 & 465 & 0 		& 85.2 & 2.2 & 69.6\\
		100 & 1200  & 164 	& 166.8 & 3.00 & 455 & 0.10 	& 85.7 & 1.5 & 69.8\\
		50 	& 1210 	& 156 	& 167.0 & 2.85 & 443 & 0.11 	& 85.8 & 1.1 & 60.5\\
		12 	& 1250 	& 169 	& 166.9 & 2.36 & 496 & 0.18 	& 85.9 & 1.1 & 71.5\\
		6 	& 1260 	& 183  	& 167.1 & 2.50 & 494 & 0.20   	& 86.0 & 1.0 & 88.3\\
		\hline
		\end {tabular}
		\end {table*}

	Two prominent phonon absorptions at $\sim 86 cm^{-1}$ (labeled as $\alpha$) and $\sim 167 cm^{-1}$ (labeled as $\beta$) are seen in $R(\omega)$ and \s~at all temperatures. A simple group symmetry analysis\cite{Kroumova2003} indicates that the \sample~should have four infrared(IR)-active modes of [2$A_{2u}$ + 2$E_{u}$] and two $E_{u}$ modes can be observed by in-plane optical measurement. Indeed, this is the case in our in-plane infrared measurement. As temperature decreases, the $\alpha-$phonon shifts very slightly to higher energy and becomes sharper. Meantime, the $\beta$-phonon's line shape shows a strong asymmetry at low temperature, which results in a dip structure just above the phonon's center-frequency. This is a typical Fano lineshape resulting from the strong coupling of phonon mode with the electronic continuum of excitations \cite{Fano1961}. The asymmetric Fano lineshape of phonons is also observed in many other compounds such as $\mathrm{Bi_2Se_3}$\cite{LaForge2010}, FeSi\cite{Damascelli1997,Krannich2015}, and even graphene\cite{Kuzmenko2009,Tang2010}.

	To better and quantitatively understand the above results we decompose the optical conductivity spectra into different components using a Drude-Lorentz analysis. The dielectric function has the form \cite{Hu2008}
	\begin{equation}
	\epsilon(\omega)=\epsilon_{\infty}-\frac{\omega_{p}^2}{\omega^2+i\omega/\tau} +\sum\limits_{j} \frac{\Omega_{j}^2}{\omega_{j}^2-\omega^2-i\omega/\tau_j},
	\end{equation}
where $\epsilon_\infty $ is the dielectric constant at high energy, and the middle and last term are the Drude and Lorentz components. The real part of conductivity for each Drude and Lorentz component has, respectively, the formula,
	\begin{equation}\label{Drude}
	\sigma_{Drude}(\omega)={\omega^2_{p} \over 4\pi\Gamma } {{\Gamma^2} \over {\omega^2+\Gamma^2}},
	\end{equation}
and
	\begin{equation}\label{lorentz}
	\sigma_{Lorentz}(\omega)={\Omega_{j}^2 \over 4\pi\Gamma_j } {{\Gamma_j^2\omega^2} \over {(\omega_j^2-\omega^2)^2+\omega^2\Gamma_j^2}}.
	\end{equation}	
Usually, a phonon can also be fit with the Lorentz model, such as the $\alpha-$phonon. However, the line shape of the $\beta$-phonon is strongly asymmetric at low temperature and cannot be fit well with the Lorentz model. Then, we employ an asymmetric Fano-lineshape model \cite{Fano1961,LaForge2010},
	\begin{equation}\label{fano}
		\sigma_{Fano}(\omega)={\omega^2_{p,Fano} \over 4\pi\Gamma } {{q^2+2q\epsilon-1} \over {q^2(1+\epsilon^2)}},
	\end{equation}
	as a function of reduced energy $\epsilon=(\omega-\omega_0)/\Gamma$. The added parameter q=-1/tan$(\Theta/2)$ determines the asymmetry of lineshape and can be considered as a measure of the degree of electron-phonon coupling. When $\Theta$ is larger, the coupling is stronger. By contrast the Lorentz line shape is recovered when $\Theta=0$ or equivalently,  $|q|\rightarrow \infty$. The sign of $\Theta$ sets the direction of asymmetry. Therefore, the  total real part of conductivity can be written as
	\begin{equation}\label{model}
		  \sigma_{1}(\omega) = \sigma_{Drude}(\omega)+\sigma_{Fano}(\omega)+\sum_{i}{ \sigma_{Lorentz,i}(\omega)},
	\end{equation}
where the summation index \emph{i} refers to the \emph{i}th Lorentz component if more than one Lorentz components are employed.

	As shown in Fig. \ref{fig:fit}, the main phonon peaks (insets) and conductivity spectrum at 300 K can be well reproduced by Eq. (\ref*{model}). Besides a Drude component and two phonon modes at low frequencies, extra Lorentz components are added to reflect the high energy interband transitions. In Fig. \ref{fig:fit}, a temperature-independent Lorentz component with parameters of $\omega_1$=2600 \cm, $\Omega_1$=5700 \cm and $\Gamma_1$=6800 \cm is also displayed. Its central frequency is beyond the frequency range of the horizontal axis. While at T = 6 K the spectrum below 60 \cm~only crudely matches with the fit results because of the invalidation of Drude model due to the too strong magnetic scattering. The fit parameters for the Drude component and two phonon modes are presented in Table. \ref{FitTable}. We find that the plasma frequency \omegap~ is only around 1000 \cm. The very small value yields clear evidence that the compound is a very low carrier density metal. Taking the charge neutrality into account, the compound should be a low carrier density semimetal.

It deserves to remark that the asymmetric $\beta$-phonon mode shows interesting temperature evolution. At high temperature above 100 K, the phonon mode is almost symmetric and the phonon peak can be fit with a Lorentz lineshape or Fano model with $\Theta$ value being set to zero. At low temperature when the scattering between free carriers and local moments becomes strong, the $\Theta$ value is non-zero and becomes larger. The data seem to imply that the enhancement of electron-lattice coupling is also related to the strength of magnetic fluctuation or ordering of the compound. Our measurements suggest that $\mathrm{EuCd_2As_2}$ is a promising candidate for exploration of intriguing interplay among charge, magnetism and the underlying crystal lattice.

Finally we wish to comment that the layered triangle lattices, together with associated electronic and magnetic properties, are of high interest in the condensed matter community. The spin-orbit-coupled rare earth triangular lattice magnetic systems have recently attracted much attention. For example, the ytterbium-based layered compound YbMgGaO$_4$, where the rare earth magnetic ions Yb$^{3+}$ form a perfect triangular lattice, was found to be disordered down to the lowest measurable temperature 60 mK \cite{Li2015}. It was suggested to be a U(1) quantum spin liquid with a spinon Fermi surface \cite{Li2015,Li2015b}. The layered family ReCd$_3$Pn$_3$ (Re=Ce and rare earth elements, Pn=P, As) with rare earth elements Re also forming a triangular lattice were found to host either disordered spin liquid or antiferromagnetic orders at different temperatures \cite{Higuchi2016a,Li2015b}. The ReCd$_3$Pn$_3$ family has the space group P63/mmc, which resembles in many aspects to the crystal structure studied in the present work. It is expected that by choosing different rare earth ions at the Ca site and other elements at Al and Si sites in the $\mathrm{CaAl_2Si_2}$-type structure, more interesting magnetic and electronic properties could be found in this structural family. We hope that the present work could inspire further studies on this and related compounds in this structure family.

\section{Conclusion}
	To conclude, we present anisotropic transport and optical spectroscopy studies on $\mathrm{EuCd_2As_2}$. The compound shows metallic transport property with moderate anisotropic ratio. At low temperature, the conducting carriers suffer strong scattering from the Eu magnetic moments, resulting in a Kondo-like increase of resistivity. Below $T_N$, the resistivity drops sharply due to the reduced scattering from the ordered Eu moments. However, the anisotropic ratio of $\rho_c/\rho_{ab}$ keeps increasing, suggesting that the antiferromagnetic coupling is along the c-axis. The optical spectroscopy measurement reveals that $\mathrm{EuCd_2As_2}$ has a small plasma frequency with an overdamped shape of reflectance edge, yielding further evidence that the compound is a low carrier density semimetal with strong carrier scattering effect. Two infrared-active phonons superimposed in the electronic background are observed in the far-infrared spectra. One of them shows a typical Fano line-shape, indicating a strong coupling between phonon and electronic continuum. The work establishes $\mathrm{EuCd_2As_2}$ as an interesting material displaying coupling among different degrees of freedoms. We hope this work will stimulate further studies on rare earth element based triangular lattice magnetic systems.

\begin{acknowledgments}
	We acknowledge useful help and discussions with T. Dong, J. L. Luo, P. Zheng, H. M. Weng and G. Chen. This work is supported by the National Science Foundation of China ("Grants No. 11120101003, 11327806,11274367,11474330), and the 973 project of the Ministry of Science and Technology of China ("Grants No. 2011CB921701, 2012CB821403).
\end{acknowledgments}

\bibliographystyle{apsrev4-1}
\bibliography{EuCd2As2}

\end{document}